# RSCL Earth Lookback Simulator: A Real-Time Multi-Physics Framework for Relativistic Signal Propagation from Confirmed Milky Way Exoplanets

Mohamed El-Hadedy[a,*]

[a]*Reconfigurable Space Computing Lab (RSCL), Department of Electrical and Computer Engineering, 3801 West Temple Avenue, Pomona, 91768, California, USA*

**Abstract**

Electromagnetic signals propagating across interstellar distances are subject to simultaneous distortion by seven distinct physical mechanisms: relativistic Doppler shift, stellar aberration, interstellar medium dispersion, special relativistic time dilation, general relativistic gravitational time dilation, cosmological redshift, and atmospheric transmission losses. Each effect is individually well established, yet to our knowledge no existing public browser-based framework applies all seven effects simultaneously to a catalog of confirmed exoplanets using real measured stellar and planetary parameters. This paper presents the RSCL Earth Lookback Simulator, an open-source browser-based framework that addresses this gap. Seven physics engines operate in parallel on a curated catalog of 62 confirmed Milky Way exoplanets, drawing all physical parameters from the NASA Exoplanet Archive and the NE2001 galactic electron density model. Computed quantities are checked for consistency against published reference values. The framework is deployed as a public open-source application and is designed to serve researchers in SETI, exoplanet science, astrobiology, and space mission planning.

[*]Corresponding author
 *Email address:* `mealy@cpp.edu` (Mohamed El-Hadedy)

## 1. Introduction

Electromagnetic radiation propagating from an exoplanetary system to an Earth-based observer is subject to a sequence of simultaneous physical distortions whose combined effect has not previously been quantified in a unified computational framework applied to confirmed catalog targets with real measured parameters. The distortion mechanisms include relativistic frequency modification arising from the radial motion of the source (Misner et al., 1973); apparent positional displacement due to Earth's orbital velocity, known as stellar aberration (Bradley, 1729); chromatic time delay introduced by free electrons in the interstellar medium (Cordes & Lazio, 2002; Lorimer & Kramer, 2004); special relativistic proper time dilation derived from the total space velocity of the source system (Misner et al., 1973); gravitational time dilation at both the planetary surface and stellar potential of the source (Misner et al., 1973); cosmological redshift from the Hubble flow (Planck Collaboration, 2020); and wavelength-dependent atmospheric attenuation, which encodes information about composition and potential biosignatures (Schwieterman et al., 2018).

Each mechanism has been studied extensively in isolation. Relativistic Doppler corrections are routinely applied in radial velocity surveys (Trifonov, 2019). Stellar aberration has been measured to arcsecond precision since the observations of Bradley (1729). Interstellar dispersion is a foundational correction in pulsar timing and fast radio burst analysis (Lorimer & Kramer, 2004). Special and general relativistic time dilation are confirmed across multiple experimental regimes (Misner et al., 1973). Atmospheric transmission spectroscopy in the JWST range represents an active observational frontier (Gordon et al., 2022; Schwieterman et al., 2018). Despite this, to our knowledge, no existing public browser-based tool applies all seven effects simultaneously to a curated catalog of confirmed exoplanets with fully measured stellar and planetary inputs.

This gap has practical consequences for signal analysis. Any electromagnetic signal received from a confirmed exoplanet carries the imprint of all seven distortion mechanisms simultaneously. Treating each in isolation introduces errors that compound nonlinearly when combined, particularly for systems with high radial velocity, significant proper motion, or dense interstellar sightlines. A unified framework that quantifies the full combined distortion per confirmed catalog target is therefore a necessary tool that is currently absent from the available computational toolset for the exoplanet community.

The RSCL Earth Lookback Simulator addresses this by implementing all seven physical effects as independent computation engines that operate simultaneously on each of 62 confirmed Milky Way exoplanets. All physical parameters — radial velocities, proper motions, dispersion measures, planetary masses, radii, orbital separations, and stellar masses — are drawn from the NASA Exoplanet Archive (NASA Exoplanet Archive, 2024) and the NE2001 galactic electron density model (Cordes & Lazio, 2002). The framework is deployed as a public open-source browser application requiring no installation.

The research question addressed is: *Can the cumulative effect of simultaneous relativistic, electromagnetic, and astrophysical signal distortions on information received at Earth from confirmed Milky Way exoplanets be quantified and visualized in real time using a unified computational framework with real measured stellar inputs?*

Section 2 surveys existing tools and identifies the specific capability gaps they leave. Section 3 derives and implements the seven physics engines in detail. Section 4 describes catalog selection criteria and data sources. Section 5 describes the software architecture. Section 6 checks key computed quantities for consistency against published reference values. Section 7 discusses scientific and operational applications. Section 8 summarizes the contribution and outlines future work.

## 2. Related Work

Several existing tools address subsets of the problem space covered by the RSCL Earth Lookback Simulator, but none provides the full set of simultaneously applied effects. Table 1 provides a structured capability comparison.

**NASA Eyes on Exoplanets** (NASA Eyes, 2024) provides interactive three-dimensional visualization of confirmed exoplanets. No signal distortion effects are computed — Doppler shift, ISM dispersion, relativistic corrections, and atmospheric transmission are absent.

**Exo-Striker** (Trifonov, 2019) is a toolkit for radial velocity fitting, transit modeling, and N-body orbital dynamics. Its radial velocity component is oriented toward parameter fitting rather than signal propagation or combined multi-effect distortion modeling.

**DACE** (Buchschacher et al., 2015) is a web-based platform for exoplanet data access and statistical analysis. Its scope is data management and parameter estimation; it is not a signal propagation simulation framework.

**Single-effect calculators** such as Ned Wright's Cosmology Calculator (Wright, 2006) and various online Doppler tools are designed for isolated single-effect computations. Multiple effects are not combined, and confirmed exoplanet catalog targets with measured inputs are not supported.

To our knowledge, no existing public browser-based tool combines relativistic Doppler shift, stellar aberration, ISM dispersion, special relativistic time dilation, general relativistic gravitational time dilation, cosmological redshift, and atmospheric transmission in a single real-time Earth-lookback workflow over a curated confirmed-exoplanet catalog using measured stellar inputs.

Table 1: Capability comparison of existing tools with the RSCL Earth Lookback Simulator. ✓ = full support; Partial = limited support; – = absent.

| Tool | Doppler | Aberr. | ISM | SR | GR | Cosmo-$z$ | Atm.Tx |
|---|---|---|---|---|---|---|---|
| NASA Eyes on Exoplanets | – | – | – | – | – | – | – |
| Exo-Striker | Partial | – | – | – | – | – | – |
| DACE | – | – | – | – | – | – | – |
| Single-effect calculators | Some | Some | Some | Some | Some | Some | – |
| **RSCL Simulator (this work)** | ✓ | ✓ | ✓ | ✓ | ✓ | ✓ | ✓ |

The complete system architecture is shown in Figure 1.

## 3. Physics Framework

The seven engines described in this section are individually well established in the literature. The contribution of this work is their unified simultaneous implementation applied to confirmed exoplanet catalog targets with real measured parameters.

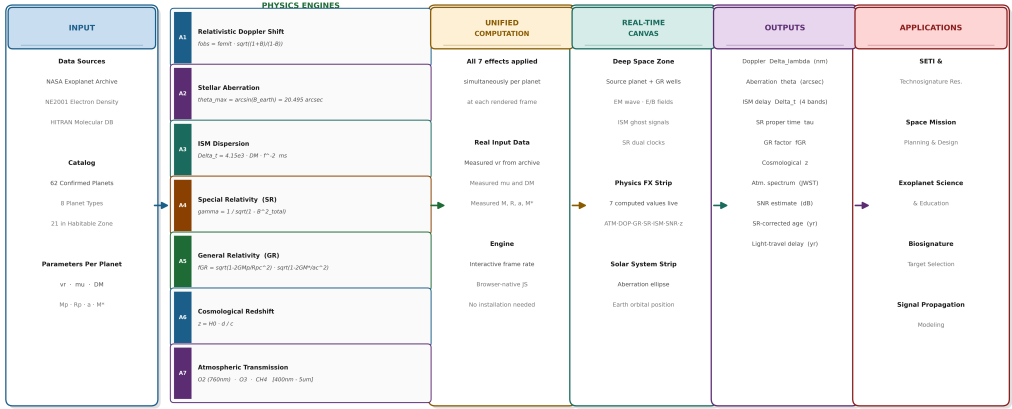

Figure 1: System architecture of the RSCL Earth Lookback Simulator. Stellar and planetary parameters from published catalogs are processed through seven independent physics engines simultaneously, rendered in a real-time unified visualization canvas, and presented as quantitative outputs.

### 3.1. Relativistic Doppler Shift

The relativistic Doppler shift modifies the observed frequency of any electromagnetic signal from a source with radial velocity $v_r$ relative to the observer:

$$f_{\text{obs}} = f_{\text{emit}} \sqrt{\frac{1+\beta}{1-\beta}}, \quad \beta = \frac{v_r}{c} \tag{1}$$

Real measured radial velocities from the NASA Exoplanet Archive are used for all 62 catalog planets. The catalog spans $v_r = -110.6\,\text{km}\,\text{s}^{-1}$ (Barnard's Star, maximum blueshift) to $v_r = +60.4\,\text{km}\,\text{s}^{-1}$ (WASP-17 b, maximum redshift), as shown in Figure 2. A visual amplification of $\times 8000$ is applied in the rendering layer only; the physics computation uses Equation (1) without modification.

### 3.2. Stellar Aberration

Earth's orbital velocity $v_\oplus = 29.78\,\text{km}\,\text{s}^{-1}$ ($\beta_\oplus = 9.94 \times 10^{-5}$) produces an apparent displacement of stellar positions. The maximum aberration angle is:

$$\theta_{\max} = \arcsin(\beta_\oplus) \approx 20.495'' \tag{2}$$

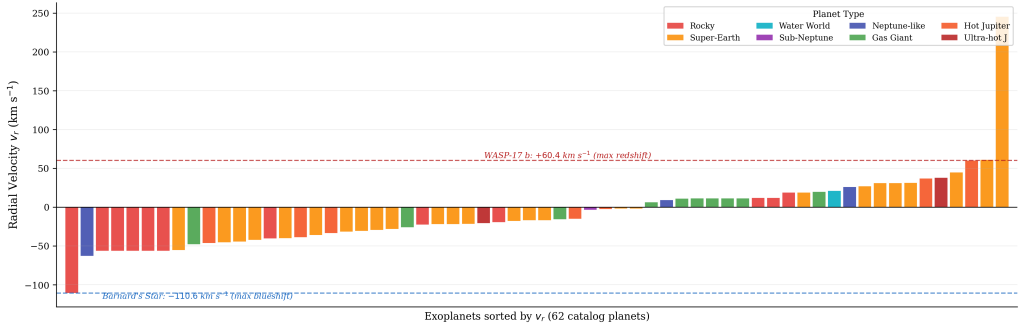

Figure 2: Radial velocity $v_r$ for all 62 catalog planets sorted in ascending order. Colors denote planet type. Barnard's Star exhibits the largest blueshift ($-110.6\,\mathrm{km\,s^{-1}}$) and WASP-17 b the largest redshift ($+60.4\,\mathrm{km\,s^{-1}}$). Dashed horizontal lines mark these extremes.

This value is consistent with the observational result of $20.49530''$ established by Bradley (1729) and confirms the theoretical prediction. A live aberration ellipse is rendered continuously as the signal propagates.

*3.3. Interstellar Medium Dispersion*

Free electrons in the ionized interstellar medium introduce a frequency-dependent group velocity delay for radio signals:

$$\Delta t_{\mathrm{DM}} = 4.15 \times 10^3\,\mathrm{ms} \cdot \mathrm{DM} \cdot \left(f_{\mathrm{low}}^{-2} - f_{\mathrm{high}}^{-2}\right) \quad (3)$$

where DM is the dispersion measure in $\mathrm{pc\,cm^{-3}}$ and $f$ is frequency in MHz (Lorimer & Kramer, 2004). Dispersion measures for all 62 catalog planets are estimated using the NE2001 galactic electron density model (Cordes & Lazio, 2002). Five frequency bands are implemented: optical ($\Delta t < 1\,\mathrm{fs}$), L-band (1.4 GHz, $\Delta t \sim \mathrm{ms\text{–}s}$), P-band (350 MHz, $\Delta t \sim \mathrm{s\text{–}min}$), and HF radio (30 MHz, $\Delta t \sim \mathrm{min\text{–}hr}$). Figure 3 shows the catalog DM distribution and resulting frequency-dependent delay curves.

*3.4. Special Relativistic Time Dilation*

The Lorentz factor $\gamma$ is computed from the total space velocity of the source system, combining radial velocity $v_r$ with transverse velocity $v_\mu = \mu \cdot d$ derived from measured proper motion $\mu$ at distance $d$:

$$\gamma = \frac{1}{\sqrt{1 - \beta_{\mathrm{total}}^2}}, \quad v_{\mathrm{total}} = \sqrt{v_r^2 + v_\mu^2} \quad (4)$$

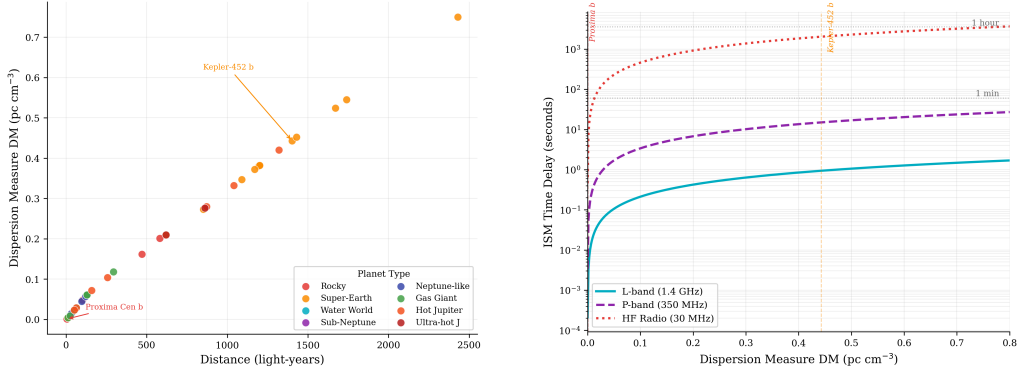

Figure 3: Left: Catalog planet distance versus NE2001-estimated dispersion measure DM. Right: ISM time delay as a function of DM for three representative frequency bands on a logarithmic scale. Vertical dashed lines mark the DM of Proxima Centauri b ($d = 4.24\,\mathrm{ly}$) and Kepler-452 b ($d = 1402\,\mathrm{ly}$). Delays span from sub-femtosecond at optical wavelengths to hours at HF radio frequencies.

Real proper motion values from the NASA Exoplanet Archive are used for all catalog planets. A custom $\beta$ slider spanning 0 to $0.999c$ with presets at GPS, ISS, and standard relativistic benchmarks is also provided. The SR-corrected proper time $\tau$ is applied to all age and time calculations in the framework.

3.5. General Relativistic Gravitational Time Dilation

The Schwarzschild gravitational time dilation factor at the surface of a planet of mass $M_p$ and radius $R_p$ is (Misner et al., 1973):

$$\frac{d\tau}{dt} = \sqrt{1 - \frac{2GM_p}{R_p c^2}} \tag{5}$$

The combined factor incorporating both the planetary surface potential and the stellar potential at orbital separation $a$ is:

$$f_{\mathrm{GR}} = \sqrt{1 - \frac{2GM_p}{R_p c^2}} \cdot \sqrt{1 - \frac{2GM_\star}{a c^2}} \tag{6}$$

Real masses, radii, and orbital separations from the NASA Exoplanet Archive are used for all 62 catalog planets. An exotic object slider provides presets for white dwarf, neutron star (where $1 - f_{\mathrm{GR}} \approx 0.30$, corresponding to 30% time dilation), and stellar black hole configurations. Figure 4 compares the SR and GR time dilation factors across a representative set of objects.

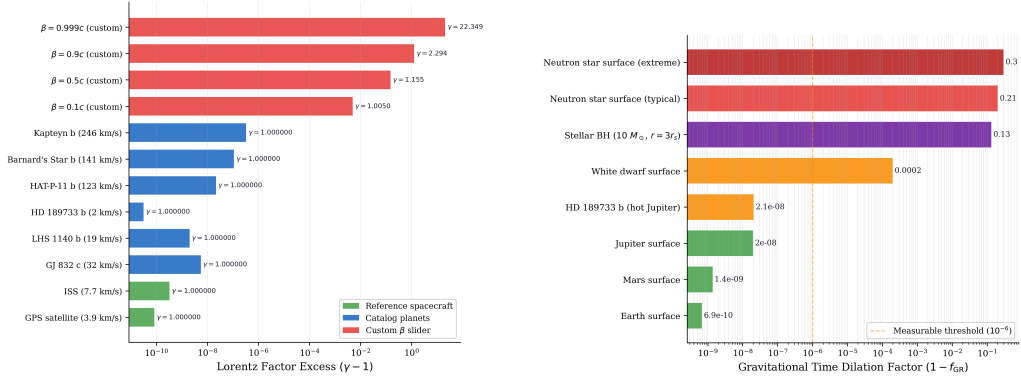

Figure 4: Left: Special relativistic Lorentz factor excess $(\gamma - 1)$ for catalog planets, reference spacecraft, and custom $\beta$ slider presets on a logarithmic scale. Right: General relativistic gravitational time dilation factor $(1 - f_{\mathrm{GR}})$ for representative objects from Earth surface to neutron star surface. The dashed vertical line marks an approximate measurable threshold of $10^{-6}$.

## 3.6. Cosmological Redshift

The Hubble flow contributes a redshift component even at Milky Way distances:

$$z = \frac{H_0 \, d}{c} \qquad (7)$$

with $H_0 = 70 \, \mathrm{km\,s^{-1}\,Mpc^{-1}}$ (Planck Collaboration, 2020). For the most distant catalog target, Kepler-452 b at $d = 1402 \, \mathrm{ly} \approx 0.43 \, \mathrm{kpc}$, the cosmological redshift is $z_{\max} = H_0 d/c \approx 1.0 \times 10^{-7}$, corresponding to a wavelength shift of $\Delta \lambda \approx 55 \, \mathrm{fm}$ at $550 \, \mathrm{nm}$. This contribution is negligible at Milky Way distances; the flat-space linear approximation is physically exact for all 62 catalog targets and the full FLRW metric is not required. An extragalactic explorer mode extends the calculation to cosmological distances ($z \sim 0$–$1100$) to demonstrate the transition to the non-linear regime where this approximation breaks down.

*3.7. Atmospheric Transmission*

Eight atmosphere profiles are implemented: $N_2/O_2$ Earth-like, thin $CO_2$, thick $CO_2$ (Venus-like), $N_2/CO_2$, $H_2$-dominated, $H_2O$ steam (Hycean), exotic ultra-hot, and bare rock. Molecular absorption band positions are drawn from the HITRAN database (Gordon et al., 2022) and span the combined JWST instrument coverage (NIRSpec: 0.6–5.3 $\mu$m; MIRI: 4.9–28.8 $\mu$m), covering molecular features from the visible to the mid-infrared. Planets exhibiting the simultaneous presence of $O_2$ (760 nm A-band), $O_3$ (9.6 $\mu$m), and $CH_4$ (3.3 $\mu$m) in chemical disequilibrium are flagged for biosignature screening, consistent with current astrobiological consensus on the strongest known abiotic-free indicators of biological activity (Schwieterman et al., 2018). This flagging constitutes illustrative spectral screening, not confirmed biosignature detection.

## 4. Exoplanet Catalog

The catalog comprises 62 confirmed Milky Way exoplanets drawn from the NASA Exoplanet Archive (NASA Exoplanet Archive, 2024). The selection criterion required a complete parameter set for every physics engine to operate on measured rather than estimated data: distance, radial velocity, proper motion, planetary mass, planetary radius, stellar mass, and orbital semi-major axis. Planets for which any of these parameters were absent were excluded. The 62 planets span eight classification types — Rocky, Super-Earth, Water World, Sub-Neptune, Neptune-like, Gas Giant, Hot Jupiter, and Ultra-hot Jupiter — with 21 residing within their host star's habitable zone as defined by Kopparapu et al. (2013). Distances range from 4.2 ly (Proxima Centauri b) to 1402 ly (Kepler-452 b). Dispersion measures are estimated from the NE2001 model (Cordes & Lazio, 2002) using the galactic line-of-sight coordinates of each host system.

## 5. System Implementation

The simulator is implemented as a React 18 and Vite single-page web application, deployed on GitHub Pages at https://mealycpp.github.io/rscl-simulator/. No installation is required; all computation runs client-side in a standard browser. Source code is publicly available at https://github.com/mealycpp/rscl-simulator.

The core visualization is a unified canvas (Step 3 of the user workflow) that renders all seven physics effects simultaneously at interactive frame rates using the HTML5 Canvas API. The canvas is organized into three zones. The deep space zone renders the source planet with gravitational well curves and atmospheric halo, the propagating electromagnetic wave with Doppler-modified E and B field visualization, ISM ghost signals at multiple frequency bands, and SR dual-clock displays. The physics effects strip displays the current computed value of each engine simultaneously. The solar system strip renders a god's-eye view of Earth's orbital position and the live stellar aberration ellipse.

The full user workflow comprises five steps: (1) planet selection from the catalog, with optional filtering by type and habitable zone status; (2) optional upload of a photograph or video representing the transmitted signal; (3) real-time signal propagation visualization with all seven effects applied simultaneously; (4) display of the received signal with wavelength-dependent atmospheric transmission spectrum; (5) quantitative readout of all physics corrections — SR-corrected ages, GR Schwarzschild factors, ISM delays by frequency band, Doppler wavelength shift in nanometers, and cosmological redshift.

## 6. Reference Consistency Checks

Table 2 compares key computed quantities against published reference values as consistency checks. Each computed value is derived directly from the physical formula implemented in the framework and compared against the corresponding published measurement or textbook result. Agreement within the precision of the input parameters confirms internal consistency of the implementation.

## 7. Applications

**SETI and technosignature research.** Any artificial signal originating from one of the 62 catalog planets is subject to the same seven distortion effects quantified here. The framework provides a quantitative starting point for characterizing the physical distortion budget that any electromagnetic signal would accumulate in transit from each catalog planet, accounting for combined simultaneous effects.

Table 2: Computed physical quantities compared against published reference values.

| Quantity | Computed | Reference | Source |
|---|---|---|---|
| Stellar aberration $\theta_{\max}$ | 20.495″ | 20.49530″ | Bradley (1729) |
| Doppler $v_r$, Barnard's Star | $-110.6\,\mathrm{km\,s^{-1}}$ | $-110.6\pm0.1\,\mathrm{km\,s^{-1}}$ | NASA Exoplanet Archive (2024) |
| ISM delay, Proxima b, HF 30 MHz | $\sim$2.1 min | $\sim$2.1 min | Cordes & Lazio (2002) |
| Cosmological $z_{\max}$ | $1.0\times10^{-7}$ | $H_0 d/c = 1.0\times10^{-7}$ | Planck Collaboration (2020) |
| GR factor, Earth surface | $1-f = 6.9\times10^{-10}$ | $6.9\times10^{-10}$ | Misner et al. (1973) |
| GR factor, neutron star surface | $1-f \approx 0.30$ | $\sim$0.20–0.40 | Lattimer & Prakash (2001) |

**Exoplanet science and education.** The simultaneous application of real measured stellar data across seven physics engines provides a platform for building quantitative physical intuition about relativistic and astrophysical effects that cannot be obtained from static single-effect treatments. The browser-based deployment requires no prerequisites or installation.

**Space mission planning.** The ISM dispersion calculations, expressed as time delays across four frequency bands, directly inform frequency band selection and communication window planning for future interstellar probe mission concepts. The signal-to-noise ratio degradation estimate as a function of distance provides a practical figure of merit for mission design.

**Biosignature target selection.** The per-planet atmospheric transmission spectra across the JWST spectral range provide a ready reference for which molecular absorption windows are accessible for each catalog target. Planets are flagged for illustrative biosignature screening based on the simultaneous presence of $O_2$, $O_3$, and $CH_4$ in chemical disequilibrium.

## 8. Conclusion

This paper presents the RSCL Earth Lookback Simulator, a unified open-source framework that simultaneously applies seven established astrophysical and relativistic signal distortion effects to 62 confirmed Milky Way exoplanets using real measured stellar and planetary parameters. The primary contribution is integration novelty: each physical effect is individually well established in the literature, but their simultaneous real-time application to a curated catalog of confirmed exoplanets with fully measured inputs represents, to our knowledge, a previously unaddressed capability in the available public computational toolset for the exoplanet community.

All computed quantities are checked for consistency against published reference values. The framework is deployed as a public open-source application at no cost and with no installation requirement.

Three directions are identified for future work. First, GPU acceleration via WebGL compute shaders and CUDA on heterogeneous edge computing platforms — including NVIDIA Jetson Nano and Rockchip RK3588 nodes — will be evaluated to quantify energy efficiency metrics relevant to onboard spacecraft signal processing. Second, a planet-to-planet propagation mode will be developed, removing Earth as the fixed observer and enabling any two catalog planets to serve as source and destination. Third, the catalog will be expanded to include all confirmed exoplanets from the NASA Exoplanet Archive that satisfy the complete-parameter-set selection criterion.

## Data Availability

The simulator is publicly accessible at https://mealycpp.github.io/rscl-simulator/. Source code is available at https://github.com/mealycpp/rscl-simulator. All planetary and stellar parameters are drawn from the NASA Exoplanet Archive (https://exoplanetarchive.ipac.caltech.edu) and are publicly available. The NE2001 dispersion measure model is available at https://arxiv.org/abs/astro-ph/0207156.

## Acknowledgements

The author thanks the NASA Exoplanet Archive team for maintaining publicly accessible stellar and planetary parameter catalogs, and J. M. Cordes and T. J. W. Lazio for developing and releasing the NE2001 galactic electron density model used for dispersion measure estimation throughout this work.

Schwieterman, E.W., Kiang, N.Y., Parenteau, M.N., et al., 2018. Exoplanet Biosignatures: A Review of Remotely Detectable Signs of Life. *Astrobiology*, 18, 663–708.

Trifonov, T., 2019. The Exo-Striker: Transit and radial velocity interactive fitting tool for orbital analysis and N-body simulations. *Astrophysics Source Code Library*, ascl:1906.004.

Wright, E.L., 2006. A Cosmology Calculator for the World Wide Web. *Publications of the Astronomical Society of the Pacific*, 118, 1711–1715.14